\documentstyle[12pt]{article}
\normalbaselines
\begin{document}

\baselineskip=24pt

\bibliographystyle{unsrt}
\vbox{\vspace{6mm}}

\begin{center}
{\Large Relativistic properties of  ``marginal''
distributions} 
\end{center}

\bigskip 

\begin{center}
Stefano Mancini$^{\dag}$, Vladimir I. Man'ko$^{\ddag}$ 
and Paolo Tombesi$^{\dag}$
\end{center}

\bigskip

\begin{center}
{\it
$\dag$Dipartimento di Matematica e Fisica and Unit\'a INFM,\\ 
Universit\`a di Camerino,
I-62032 Camerino, Italy \\
$\ddag$P.N. Lebedev Physical Institute, Leninsky Prospekt 53,
Moscow 117934, Russia}
\end{center}

\bigskip
\bigskip
\bigskip

\begin{abstract}
We study the properties of marginal distributions-projections of 
the phase space representation of
a physical system-under relativistic transforms.
We consider the Galileo case as well as the Lorentz transforms 
exploiting
the relativistic oscillator model used for describing the
mass spectrum of elementary particles.
\end{abstract}

PACS number(s): 03.65.Bz, 03.30.+p, 03.65.Ca

\bigskip

\section{Introduction}

The concept of phase space arises naturally from the Hamiltonian 
formulation of classical
mechanics, and there have always been considerable efforts to 
give phase space picture
of quantum mechanics too.  Much of the thrust of these attempts 
lies in their ability to exploit
classical analogues. Using these techniques, such as 
$P$-representation of Glauber and Sudarshan
\cite{glasud}, the Wigner representation \cite{wig} and the 
Husimi representation \cite{hus}, some
quantum systems can be reduced to non-operator systems. 
However, the essential quantum nature of
the problem is present in terms of the interpretation of 
the (apparently) classical variables.

Moreover, the so called quasi-probability distributions 
\cite{cahgla} do not have the 
characteristics
of classical probability distributions. 
Instead, by projecting the quasi-probability in a certain 
phase subspace, it is possible to
obtain a genuine probability.
In particular the projection of the Wigner function onto a 
straight line of the phase space was
called `marginalization' procedure, and the obtained 
distribution `marginal' \cite{wig,ber}.

Recentely, there has been a renewed interest on these  
marginal probabilities in connection with 
the tomographic imaging of a quantum state \cite{ray}.
Along this approach the marginals represent the shadows 
from which the state (or its phase space
representation) is reconstructed \cite{vr}.

The aim of the present paper is to study the properties 
of these marginals under relativistic
transforms. The case of Galileo transforms results almost 
trivial, while for the Lorentz one
we needed of a model having a covariant phase space picture. 
To this end we have studied the
relativistic oscillator model used for describing the mass 
spectrum of elementary particles
\cite{fey}.

\section{The ``marginal" distributions}

Referring to the standard definitions given in the 
literature \cite{wig,ber}, by
`marginalization' one should mean a line integral in 
the phase spce $\{q,p\}$ of the Wigner function
$W(q,p)$, i.e.
\begin{equation}\label{standdef}
w(x;\theta)=\int\,dq\,dp\,W(q,p)\,\delta(x-\cos\theta 
q-\sin\theta p)\,,
\end{equation}
where $\theta$ is the angle orientation of the line.
$w$ becomes a probability distribution for the variable $x$, 
depending parametrically on $\theta$.

One can go beyond this definition \cite{qso97}; let us 
consider the phase space transformation
as a generic linear combination of position
$q$ and momentum $p$
\begin{eqnarray}
q\rightarrow X&=&\mu q+\nu p\,,\label{Xtra}\\
p\rightarrow{\cal P}&=&\mu'q+\nu'p\,,\label{Ptra}
\end{eqnarray}
and consider it as a real symplectic 
transformation belonging to the 
group $Sp(2,R)$, i.e.
\begin{equation}\label{sym}
\Lambda \sigma \Lambda ^{T}=\sigma\,,\quad
\Lambda =\left (\begin{array}{clcr}
\mu &\nu \\
\mu '&\nu '\end{array}\right )\,,\quad
\sigma =\left (\begin{array}{clcr}
0&1\\
-1&0\end{array}\right ).
\end{equation}

Then, we can consider the `projection' along the phase 
subspace  
characterized by the transformation (\ref{Xtra})
\begin{equation}\label{margxW}
w(X;\mu,\nu)=\int 
W(q,p)\delta(X-\mu q-\nu p)~dq~dp\,,
\end{equation}
which can be intended as a marginal distribution too.
Of course, Eq.(\ref{standdef}) represents a particular 
case of Eq.(\ref{margxW}) whenever a mere
rotation in the phase space is considered. 

The above definition could also be extended to phase spaces 
of higher dimensions. For example,
in the case of two-dimensional system we will have
a phase space 
$\{{\vec q}\equiv(q_1,q_2),\,{\vec p}\equiv(p_1,p_2)\}$, 
hence we can intruduce the transform
\begin{equation}\label{pstra2}
({\vec X},{\vec{\cal P}})
=\Lambda\;({\vec q},{\vec p})^T
\end{equation}
where now $\Lambda$ is a $4\times 4$ real symplectic matrix, 
i.e.
\begin{equation}\label{5.2}
\Lambda\sigma\Lambda^T=\sigma;\quad\sigma =
\left (\begin{array}{clcr} 0&0&1&0
\\0&0&0&1
\\-1&0&0&0
\\ 0&-1&0&0\end{array}\right )\,.
\end{equation}
The components
$X_1,X_2,{\cal P}_1,{\cal P}_2$ are related to the 
homogeneos symplectic group 
$Sp(4,R)$. 
In particular, ${\vec X}$ has the following components
\begin{eqnarray}
&X_1&=\vec\mu{\vec q}+\vec\nu{\vec p}\,,\label{Xcomp1}\\
&X_2&=\vec\mu'{\vec q}+\vec\nu'{\vec p}\,,\label{Xcomp2}
\end{eqnarray}
with $\vec\mu=(\Lambda_{11},\Lambda_{12});
\,\vec\nu=(\Lambda_{13},
\Lambda_{14});
\,\vec\mu'=(\Lambda_{21},\Lambda_{22});
\,\vec\nu'=(\Lambda_{23},
\Lambda_{24})$.

Therefore, a marginal distribution 
$w(\vec X;\vec\mu,
\vec\nu,\vec\mu',\vec\nu')$, may be introduced as
a probability distribution for the variable
$\vec X$, with a dependence upon the parameters
characterizing the matrix
$\Lambda$
\begin{eqnarray}\label{margXvec}
w(\vec X;\vec\mu,\vec\nu,\vec\mu',\vec\nu')&=&
\int \,d{\vec q}\,d{\vec p}\,W({\vec q},{\vec p})\,
\delta(X_1-\vec\mu\vec q-\vec\nu\vec p)\nonumber\\
&\times&\delta(X_2-\vec\mu'\vec q-\vec\nu'\vec p)\,.
\end{eqnarray}

\section{Marginal distributions and relativity}

Now, in order to study the properties of marginal 
distributions under 
relativistic transformations we need to know the 
transformation 
properties of the wave function of the system.

At first it is instructive to consider the Galilei 
transforms
\begin{equation}
q'=q-vt\,;\quad p'=p-v\,;\quad t'=t\,,
\end{equation}
where we have considered a particle of unit mass and  
$v$ represents the relative velocity between the two 
reference frames.

Since in this case the wave function transforms, 
independently of 
the assumed model, as
\begin{equation}
\Psi(q,t)\to\Psi'(q,t)
=\exp\left(ivq-i\frac{v^2}{2}t\right)\Psi(q-vt,t)\,,
\end{equation}
then, the Wigner function correspondingly transforms as 
\begin{equation}\label{galiW}
W(q,p,t)\to W(q',p',t)=W(q-vt,p-v,t)\,.
\end{equation}

As consequence of (\ref{galiW}) we immediatly obtain 
\begin{equation}\label{galiw}
w(x,\mu,\nu)\to w_v(x,\mu,\nu)= w_0(x-\mu vt-\nu v,\mu,\nu)\,,
\end{equation}
from which a simple shift in the distribution, in moving from
one reference frame to the other, results.

Instead, in the case of Lorentz transformations we have to 
consider a specific model.

\subsection{The relativistic harmonic oscillator} 

Let us consider the relativistic oscillator model 
introduced by 
Feynmann {\it et al.} \cite{fey} to describe a hadron
consisting of two quarks bound togheter by a harmonic 
oscillator potential of unit strength
\begin{equation}\label{1.1}
\left\{-2\left[\left(\frac{\partial}{\partial x_a^{\mu}}
\right)^2
+\left(\frac{\partial}{\partial x_b^{\mu}}\right)^2\right]
+\left(\frac{1}{16}\right)\left(x_a^{\mu}-x_b^{\mu}\right)^2
+m_0^2\right\}
\phi(x_a,x_b)=0\,,
\end{equation}
where $x_a$ and $x_b$ are space-time coordinates for the 
first and second quark respectivly (we are using natural units, 
$\hbar=c=1$).
This partial differential equation has many different 
solutions depending on the choice of
variables and boundary conditions. Here we follow the 
treatment of Ref. \cite{kimbook}.

In order to simplify the Eq. (\ref{1.1}), let us introduce 
new coordinate variables
\begin{equation}\label{1.2}
X=(x_a+x_b)/2\,;\quad x=(x_a-x_b)/2\,.
\end{equation} 
The four-vector $X$ specifies where the hadron is located 
in space-time, while the variable $x$
measures the space-time separation between the quarks. 
In terms of these variables Eq.
(\ref{1.1}) can be written as
\begin{equation}\label{1.3}
\left(\frac{\partial^2}{\partial X_{\mu}^2}-m_0^2
+\frac{1}{2}\left[\frac{\partial^2}{\partial x_{\mu}^2}
+x_{\mu}^2\right]\right)
\phi(x_a,x_b)=0\,.
\end{equation} 
This equation is separable in the $X$ and $x$ variables. Thus
\begin{equation}\label{1.4}
\phi(x_a,x_b)=f(X)\psi(x)\,,
\end{equation}
and $f(X)$ and $\psi(x)$ satisfy the following differential 
equations respectively
\begin{eqnarray}
&&\left(\frac{\partial^2}{\partial X^2_{\mu}}-m_0^2
-(\lambda+1)\right)f(X)=0\,,
\label{1.5}\\
&&\frac{1}{2}\left(-\frac{\partial^2}{\partial x^2_{\mu}}
+x^2_{\mu}\right)\psi(x)
=(\lambda+1)\psi(x)\,.\label{1.6}
\end{eqnarray}
Equation (\ref{1.5}) is  a Klein-Gordon equation, and its 
solution takes the form
\begin{equation}\label{1.7}
f(X)=\exp[\pm iP_{\mu}X^{\mu}]\,,
\end{equation} 
with
\begin{equation}
-P^2=-P_{\mu}P^{\mu}=M^2=m_0^2+(\lambda+1)\,,
\end{equation} 
where $M$ and $P$ are the mass and four-momentum of the hadron 
respectively. The eigenvalue
$\lambda$ is determined from the solution of Eq. (\ref{1.6}).

As for the four momenta of the quarks $p_a$ and $p_b$, we can 
combine them into the total
four-momentum and momentum-energy separation between the quarks
\begin{equation}\label{1.8}
P=p_a+p_b\,;\quad p=\sqrt{2}(p_a-p_b)\,.
\end{equation} 
$P$ is the hadronic four-momentum conjugate to $X$. The internal 
momentum-energy separation 
$p$ is conjugate to $x$ provided that there exist wave functions 
which can be Fourier transformed.

The four-dimensional equation (\ref{1.6}) is separable in at 
least thirty-four different
coordinate systems \cite{kalnins}. Since we are quite familiar 
with the three-dimensional harmonic
oscillator equation from nonrelativistic quantum mechanics, we 
are naturally led to consider the
separation of the space and time variables, and write the 
equation (\ref{1.6}) as
\begin{equation}\label{2.1}
\left(-\nabla^2+\frac{\partial^2}{\partial t^2}+[{\bf x}^2
-t^2]\right)\psi(x)
=(\lambda+1)\psi(x)\,.
\end{equation}
If the hadron moves along the $Z$ direction which is also 
the $z$ direction, then the hadronic
factor $f(X)$ is Lorentz-transformed in the same manner of  
a scalar field.
The Lorentz transformation of the internal coordinates from 
the laboratory frame to the hadronic
rest frame takes the form
\begin{eqnarray}
&&x'=x\,,\quad y'=y\,,\nonumber\\
&&z'=(z-\beta t)/(1-\beta^2)^{1/2}\,,\label{2.2}\\
&&t'=(t-\beta z)/(1-\beta^2)^{1/2}\,,\nonumber
\end{eqnarray}
where $\beta$ is the velocity of the hadron moving along 
the $z$ direction.
The primed quantities are the coordinate variables in the 
hadronic rest frame.
In terms of the primed variables the oscillator differential 
equation is
\begin{equation}\label{2.3}
\left(-{\nabla'}^2+\frac{\partial^2}{\partial {t'}^2}
+[{{\bf x}'}^2-{t'}^2]\right)\psi(x)
=(\lambda+1)\psi(x)\,.
\end{equation}
This form is identical to that of Eq. (\ref{2.1}), due to 
the fact that the oscillator
differential equation is Lorentz-invariant \cite{kimnoz73}.

Among many possible solutions of the above differential 
equation, let us consider the form
\begin{eqnarray}\label{2.4}
\psi_{\beta}(x)&=&\left(\frac{1}{\pi}\right)
\left(\frac{1}{2}\right)^{(a+b+n+k)/2}
\left(\frac{1}{a!b!n!k!}\right)^{1/2}
H_a(x')H_b(y')H_n(z')H_k(t')\nonumber\\
&&\times\exp\left[-\frac{1}{2}({x'}^2+{y'}^2+{z'}^2
+{t'}^2)\right]\,,
\end{eqnarray}
where $a$, $b$, $n$ and $k$ are integers, and $H_a(x')$, 
$H_b(y')\,\dots$ are the Hermite
polynomials. This wave function is normalizable, but the 
eigenvalue takes the values
\begin{equation}\label{2.5}
\lambda=(a+b+n-k)\,.
\end{equation}
Thus for  a given value of $\lambda$, there are infinitely 
many possible combinations of $a$, $b$,
$n$ and $k$. The most general solution of the oscillator 
differential equation is infinitely
degenerate \cite{yuka}.
The simplest way to avoid this problem 
(at least to render finite the degeneracy),
is to invoke the 
restriction that there  should not be time-like
oscillations in the Lorentz frame in which the hadron is 
at rest, and that the integer $k$ in Eqs.
(\ref{2.4}) and (\ref{2.5}) be zero \cite{yuka,mar}.
In doing so we are led to the question of maintaining the 
Lorentz covariance with this condition.

When the hadron moves along the $z$ axis, the $k=0$ condition 
is equivalent to
\begin{equation}\label{2.6}
\left(t'+\frac{\partial}{\partial t'}\right)\psi_{\beta}(x)=0\,.
\end{equation}
The most general form of the above condition is
\begin{equation}\label{2.7}
p_{\mu}\left(x^{\mu}+\frac{\partial}{\partial x_{\mu}}\right)
\psi_{\beta}(x)=0\,.
\end{equation}
Thus the $k=0$ condition is covariant. Once this condition 
is set, we can write the wave function
belonging to this finite set as 
\begin{eqnarray}\label{2.8}
\psi_{\beta}(x)&=&\left(\frac{1}{\pi}\right)\left(\frac{1}{2}
\right)^{(a+b+n)/2}
\left(\frac{1}{a!b!n!}\right)^{1/2}
H_a(x')H_b(y')H_n(z')\nonumber\\
&&\times\exp\left[-\frac{1}{2}({x'}^2+{y'}^2+{z'}^2
+{t'}^2)\right]\,.
\end{eqnarray}
Except for the Gaussian factor in the $t'$ variable, 
the above expression is the wave function for
the three-dimensional isotropic harmonic oscillator. 

Since the above oscillator wave functions are separable in 
the Cartesian coordinate system, and
since the transverse coordinate variables are not affected 
by the boost along the $z$ direction,
we can omit the factors depending on the $x$ and $y$ variables 
when studying their Lorentz
transformation properties. Hence, the solutions satisfying the 
subsidary condition (\ref{2.7})
take the simple form
\begin{equation}\label{2.10}
\psi^n_{\beta}(z,t)=\left(\frac{1}{\pi 2^2n!}\right)^{1/2}
H_n(z')\exp\left[-\frac{1}{2}({z'}^2+{t'}^2)\right]\,,
\end{equation}
with $\lambda=n$.
This normalizable wave function, without excitations along 
the $t'$ axis, describes the internal
space-time structure of the hadron moving along the $z$ 
direction with the velocity parameter
$\beta$. If $\beta=0$, then the wave function becomes
\begin{equation}\label{2.11}
\psi^n_{0}(x,t)=\left(\frac{1}{\pi 2^2n!}\right)^{1/2}
H_n(z)\exp\left[-\frac{1}{2}({z}^2+{t}^2)\right]\,.
\end{equation}
Thus
\begin{equation}\label{2.12}
\psi^n_{\beta}(z,t)=\psi^n_0(z',t')\,.
\end{equation}
We have therefore obtained the Lorentz-boosted wave function 
by making a passive coordinate
transformation on the $z$ and $t$ coordinate variables.

\subsection{Covariant phase space}

It is possible to construct a covariant phase space for 
the relativistic harmonic oscillator by following 
Ref. \cite{kimwig}. 
Let us consider at first the Gaussian factor of the 
wave function 
(\ref{2.11}), which practically corresponds to the 
ground state,
\begin{equation}\label{8.72}
\psi^0_0(z,t)=\left(\frac{1}{\pi}\right)^{1/2}
\exp\left(-(z^2+t^2)/2\right)\,,
\end{equation}
and introduce the light cone coordinates
\begin{equation}\label{7.45}
u=(z+t)/\sqrt{2}\,,\quad v=(z-t)/\sqrt{2}\,.
\end{equation}
The latter transform as 
\begin{equation}\label{7.46}
u'=\left(\frac{1+\beta}{1-\beta}\right)^{1/2}u\,,\quad 
v'=\left(\frac{1-\beta}{1+\beta}\right)^{1/2}v\,.
\end{equation}
It is easy to see that the product $uv$ is  
Lorentz invariant.
By using such coordinates, the wave function
(\ref{8.72}) can be rewritten as
\begin{equation}\label{8.73}
\psi^0_0(z,t)=\psi_0^0(u,v)=
\left(\frac{1}{\pi}\right)^{1/2}
\exp\left(-(u^2+v^2)/2\right)\,,
\end{equation}
and, if the system is boosted, it becomes 
\begin{equation}\label{8.74}
\psi^0_{\beta}(z,t)=
\left(\frac{1}{\pi}\right)^{1/2}
\exp\left\{-\left(\frac{1}{2}\right)
\left(\frac{1-\beta}{1+\beta}u^2
+\frac{1+\beta}{1-\beta}v^2\right)\right\}\,,
\end{equation}
Practically, it undergoes a continous deformation 
as $\beta$ increases.

Analogously to the Eq. (\ref{7.45}), we may define for the 
momentum and energy the variables
\begin{equation}\label{7.48}
p_u=(p_z-p_0)/\sqrt{2}\,,\quad p_v=(p_z+p_0)/\sqrt{2}\,,
\end{equation}
and the momentum-energy wave function will be given by
\begin{eqnarray}\label{8.75}
\phi_{\beta}^0(p_u,p_v)&=&\left(\frac{1}{2\pi}\right)
\int\psi_{\beta}^0(z,t) e^{-i(zp_z-tp_0)}dzdt\,,\\
&=&\left(\frac{1}{\pi}\right)^{1/2}
\exp\left\{-\left(\frac{1}{2}\right)
\left(\frac{1+\beta}{1-\beta}p_u^2
+\frac{1-\beta}{1+\beta}p_v^2\right)\right\}\,.
\end{eqnarray}

Hence, we deal a four dimensional phase space
$\{u,v,p_u,p_v\}$ where the Wigner function can be 
defined in a canonical way
\begin{eqnarray}\label{8.78}
W^0_{\beta}(u,p_u;v,p_v)&=&\left(\frac{1}{\pi}\right)\int
\left(\psi^0_{\beta}(u+x,v+y)\right)^*\psi^0_{\beta}(u-x,v-y)
\nonumber\\
&\times&\exp\left[2i\left(p_ux+p_vy\right)\right]dxdy\,.
\end{eqnarray}
After the evaluation of the integral we obtain
\begin{eqnarray}\label{8.79}
W^0_{\beta}(u,p_u;v,p_v)&=&\left(\frac{1}{\pi}\right)^2
\exp\left\{-\left(\frac{1}{2}\right)
\left(\frac{1-\beta}{1+\beta}u^2
+\frac{1+\beta}{1-\beta}p_u^2\right)\right\}\nonumber\\
&\times&\exp\left\{-\left(\frac{1}{2}\right)
\left(\frac{1+\beta}{1-\beta}v^2
+\frac{1-\beta}{1+\beta}p_v^2\right)\right\}\,,
\end{eqnarray}
which is manifestly covariant.

\subsection{The properties of marginal probabilities}

Having a four dimensional (covariant) phase space,
the marginal distributions 
can be defined analogously to the case of Eq.(\ref{margXvec}).
Practically, a marginal distribution will be a projection
on the plane $\{{\cal U,V}\}$
determined by the equations
\begin{eqnarray}\label{UVdef}
{\cal U}&=&\mu_1u+\mu_2p_v+\nu_1v+\nu_2p_u\,,\\
{\cal V}&=&\zeta_1u+\zeta_2p_v+\eta_1v+\eta_2p_u\,,
\end{eqnarray}
where we have $\mu_i,\nu_i,\zeta_i,
\eta_i\in{\bf R}$, $(i=1,2)$. 

Without lost of generality we do not specify 
the constraints on these parameters, 
since they will be
related to the space-time asimmetry 
in the commutation relations,
which is an hard problem to face in making the
relativistic quantum mechanics, and
goes beyond the scope of the present paper.

Then, we define
\begin{eqnarray}\label{wUV}
w_{\beta}({\cal U},{\cal V};\sigma)&=&
\int\,dudvdp_udp_v\,W_{\beta}(u,p_u;v,p_v)\nonumber\\
&\times&\delta\left({\cal U}-\mu_1u-\mu_2p_v
-\nu_1v-\nu_2p_u\right)\nonumber\\
&\times&\delta\left({\cal V}-\zeta_1u
-\zeta_2p_v-\eta_1v-\eta_2p_u\right)\,,
\end{eqnarray}
where $\sigma=\{\mu_i,\nu_i,\zeta_i,\eta_i\}$, 
$(i=1,2)$.
As limiting cases we have 
$\mu_1=\eta_1=1$, and all the other parameters equal to zero,
then
\begin{equation}\label{psiuv}
w_{\beta}({\cal U},{\cal V};\sigma)=
\left|\psi_{\beta}^0(u,v)\right|^2\,;
\end{equation}
or otherwise, for 
$\nu_2=\zeta_2=1$, and all the other parameters zero, then
\begin{equation}\label{phiuv}
w_{\beta}({\cal U},{\cal V};\sigma)=
\left|\phi_{\beta}^0(p_u,p_v)\right|^2\,.
\end{equation}

It is clear from the Eq. (\ref{wUV}) 
that the marginal distribution, with such definition, 
is not covariant,
but we may rescale the variables, due to the Wigner 
function covariance,
to have
\begin{eqnarray}\label{wtrasf}
w_{\beta}^0({\cal U},{\cal V};\sigma)&=&\int\,du'dv'dp_u'dp_v'
\,W_0^0(u',p_u';v',p_v')\nonumber\\
&\times&\delta\left({\cal U}-{\overline\mu}_1u'
-{\overline\nu}_1v'
-{\overline\nu}_2p_u'
-{\overline\mu}_2p_v'\right)\nonumber\\
&\times&\delta\left({\cal V}-{\overline\zeta}_1u'
-{\overline\eta}_1v'
-{\overline\eta}_2p_u'
-{\overline\zeta}_2p_v'\right)
\end{eqnarray}
where we take into account the invariance of the measure and 
we set
\begin{eqnarray}\label{sigbe}
{\overline\mu}_i&=&\left[
\frac{1-\beta}{1+\beta}\right]^{1/2}\mu_i\,,\;
{\overline\nu}_i=\left[\frac{1+\beta}{1-\beta}
\right]^{1/2}\nu_i\,,\\
{\overline\zeta}_i&=&\left[\frac{1-\beta}{1+\beta}
\right]^{1/2}\zeta_i\,,\;
{\overline\eta}_i=\left[\frac{1+\beta}{1-\beta}
\right]^{1/2}\eta_i\,,\quad(i=1,2)\,.\nonumber.
\end{eqnarray}

From Eq. (\ref{wtrasf}) it immediately follows 
\begin{equation}\label{wbew0}
w_{\beta}^0({\cal U},{\cal V};\sigma)=w_0^0({\cal U},
{\cal V};\sigma_{\beta})\,,
\end{equation}
where $\sigma_{\beta}$ indicates the parameters
(\ref{sigbe}).

Eq. (\ref{wbew0}) defines the transformation properties of
the marginal distributions;
practically we get that different marginals correspond to the 
same measurement, but performed in different frames.
This means that the boosts connecting several frames 
could be useful to vary the parameters characterizing the 
marginal distribution.
This interpretation is in agreement with that given in 
Ref. \cite{qso96}
for the two-mode `symplectic tomography'.

The above results can be extended to the excited states of 
the relativistic oscillator as well; the only additive factor 
one has to consider is the Hermite polynomial
multypling the Gaussian of the ground state.

Of course, the result of Galilei transforms cannot be obtained
as a limiting case of the Lorentz transforms.

\section{Conclusions}

In conclusion we have studied the properties of the marginal 
distributions under relativistic
transformations.
Since they contain all the information about the quantum 
state of a system, other probabilities
related to different observables, could be derived from them, 
as well as their properties.

Finally, the discussed properties do not concern only 
fundamental questions, but
could  become interesting in quantum optics where 
``optical mesons" enter in the reality \cite{drum}, and 
in particle physics where, hopefully, the quantum state 
tomography 
concept could be applied.
Since by repeated measurements one can build up the marginal 
probabilities, one is lead to ask
the following question:
which observables should be measured to this goal
for example in high energy processes?
This subject will be addressed in future works.

\section*{Acknowledgments}
S.M. would like to thank the Istituto Nazionale di 
Fisica Nucleare (Sezione di Perugia)
for the financial support and the Lebedev Physical 
Institute for the kind hospitality during the first stage 
of this work. V.I.M. is grateful to the Russian Foundation 
for Basic Research for the 
partial support under Project No. 17222.

\end{document}